# Magneto-electric Tuning of Pinning-Type Permanent Magnets through Atomic-Scale Engineering of Grain Boundaries

Xinglong Ye[1,✉], Fengkai Yan[2,♠], Lukas Schaefer[3], Di Wang[1,4], Holger Geßwein[5], Wu Wang[1,♪], Mohammed Reda Chellali[1], Leigh T. Stephenson[2], Konstantin Skokov[3], Oliver Gutfleisch[3], Dierk Raabe[2], Horst Hahn[1], Baptiste Gault[2,6,✉], Robert Kruk[1]

[1] Institute of Nanotechnology, Karlsruhe Institute of Technology (KIT), Eggenstein-Leopoldshafen, Germany
[2] Department of Microstructure Physics and Alloy Design, Max-Planck-Institut für Eisenforschung GmbH (MPIE), Düsseldorf, Germany.
[3] Department of Material Science, Technical University Darmstadt, Darmstadt, Germany
[4] Karlsruhe Nano Micro Facility, Karlsruhe Institute of Technology (KIT), Karlsruhe, Germany
[5] Institute for Applied Materials, Karlsruhe Institute of Technology, Eggenstein-Leopoldshafen, Germany.
[6] Department of Materials, Imperial College London, London, UK

Corresponding authors. Email: xing-long.ye@kit.edu; b.gault@mpie.de

♠now at Institute of Metal Research, Chinese Academy of Science, Shenyang, China

♪now at Department of Physics, Southern University of Science and Technology, Shenzhen, China.

**Abstract**

Pinning-type magnets maintaining high coercivity, i.e. the ability to sustain magnetization, at high temperature are at the core of thriving clean-energy technologies. Among these, $Sm_2Co_{17}$-based magnets are excellent candidates owing to their high-temperature stability. However, despite decades of efforts to optimize the intragranular microstructure, the coercivity currently only reaches 20~30% of the theoretical limits. Here, the roles of the grain-interior nanostructure and the grain boundaries in controlling coercivity are disentangled by an emerging magneto-electric approach. Through hydrogen charging/discharging by applying voltages of only ~ 1 V, the coercivity is reversibly tuned by an unprecedented value of ~ 1.3 T. In situ magneto-structural measurements and atomic-scale tracking of hydrogen atoms reveal that the segregation of hydrogen atoms at the grain boundaries, rather than the change of the crystal structure, dominates the reversible and substantial change of coercivity. Hydrogen lowers the local magnetocrystalline anisotropy and facilitates the magnetization reversal starting from the grain boundaries. Our study reveals the previously neglected critical role of grain boundaries in the conventional magnetisation-switching paradigm, suggesting a critical reconsideration of strategies to overcome the coercivity limits in permanent magnets, via for instance atomic-scale grain boundary engineering.
Consequently present work . *(200 words)*

**Key words: Magneto-electric, Permanent magnets, Hydrogen, Grain boundary**

## Introduction

Permanent magnets with the ability to maintain their magnetization, i.e. a property referred to as coercivity, at high temperatures are crucial for the flourishing clean energy technologies such as electric vehicles and wind powers[1,2]. In this regard, the pinning-type magnets, in which the coercivity arises from the pinning of magnetic domain walls at nano-precipitates within the grain,[6-8] are most promising. As an example, the $Sm_2Co_{17}$-based magnet is the only candidate for use in electric motors working above 300 °C owing to its excellent temperature stability. Its coercivity is usually believed to be controlled exclusively by domain-wall pinning because of the nano-scale cellular microstructure within the grain[9-13], while the initial demagnetization at grain boundaries is considered irrelevant. However, despite intensive efforts to optimize its intra-granular microstructure, the coercivity currently only reaches 20-30% of the theoretical anisotropy field.[1,14,15] Hence, it inevitably opens the question about the possible influence of grain boundaries during the magnetization reversal of the material.

The role of grain boundaries in pinning-type magnets may be understood if they can be modified separately from the grain interior, in conjunction with measuring the associated coercivity change. Traditional processing approaches, such as heat treatments[13], plastic deformation[16] and alloying[9,17], can dramatically change the coercivity. However, these approaches often induce irreversible modification (or destruction) of the microstructure both in grain boundaries and grain interior, obscuring the assessment of their respective impact on coercivity. Consequently, decoupling the separate roles of the grain boundary and the grain interior in magnetization reversal becomes technically challenging.

It has recently been demonstrated that the magneto-electric approach can reversibly modify the magnetic properties of materials with external voltages without changing the microstructure[8,19]. For instance, voltage-driven proton pumping and the voltage-controlled hydrogen insertion/extraction can substantially tune the coercivity of ferromagnetic metals.[20,21] Owing to different affinities of hydrogen atoms to the microstructural defects, hydrogen atoms are expected to diffuse first along the grain boundaries, and, then into the grain interior.[22,23] This sequential diffusion, if controlled, offers an opportunity to decouple the roles of the grain boundary and the grain interior, if the associated coercivity change is monitored at each step. Here, by employing electrochemically-controlled hydrogen charging/discharging, we tuned the coercivity of the $Sm_2Co_{17}$-based hard magnet by ~1.3 T, the largest values ever achieved by magneto-electric approaches. The combined in situ magneto-structural measurements and atomic-scale mapping of hydrogen distribution[24] reveal that hydrogen atoms strongly segregate at grain boundaries, which weakens the local magnetic anisotropy and accounts for the predominant change of coercivity. Our study opens a way to achieve giant magnetoelectric effects by atomic-scale engineering of grain boundaries, and unveils the critical role of grain boundaries in limiting pinning-type magnet's performance pointing the way forward for future optimisation strategies.

## Results
### Magneto-optical observation of magnetization reversal
We used commercial $Sm_2Co_{17}$-based permanent magnets with compositions of $Sm(Co_{0.766}$

Fe$_{0.116}$Cu$_{0.088}$ Zr$_{0.029}$)$_{7.35}$ (hereafter referred to as SmCo$_{7.35}$ sample) (**Table S1**). The hysteresis loop shows a coercivity of ~2.8 T (**Fig. 1A**). By magneto-optical Kerr effect (MOKE) microscopy, we observed its magnetization reversal process under demagnetization fields. Prior to MOKE imaging, the sample was fully magnetized at -6.8 T (**Fig. 1B**), and the domain structure imaged with the c-axis in the viewing plane.

At a demagnetization field of 1.5 T, the sample retained its fully-magnetized state (**Fig. 1C**). When the field increased to 2 T, the reversed domains started to appear at grain boundaries (**Fig. 1D**), and at 2.2 T expanded into the interior of the grains (**Fig.1E**). With further increasing field, the magnetic domains moved massively into the grain (**Fig. 1F**). At 3.0 T, only a few residual domains were un-reversed (**Fig. 1G**). These observations demonstrate the initial nucleation of magnetic domains at grain boundaries before their growth into the grain interior.

**Reversible modification of coercivity by non-destructive hydrogen charging/discharging**

We employed an electrochemical three-electrode cell to charge/discharge the SmCo$_{7.35}$ sample with hydrogen atoms (**Fig. S1**). In this setup, the as-prepared electrode with the SmCo$_{7.35}$ particles was the working electrode and 1 M KOH aqueous solution the electrolyte (**Fig. S2**). During hydrogen charging, the electrochemical reduction of water molecules on the metal surface provides the hydrogen ad-atoms that subsequently diffuse into the material. Conversely, during the discharging, the hydrogen atoms on the surface (H$_{ads}$) were oxidized and removed, resulting in hydrogen desorption. Based on the measured cyclic voltammogram (**Fig. S3**), we used the voltages steps of -1.2 V and -0.4 V to charge and discharge the sample, respectively.

We explored the response of the coercivity of the as-prepared SmCo$_{7.35}$ sample to hydrogen charging/discharging by *in situ* superconducting quantum interference device (SQUID). The coercivity of the as-prepared sample was ~ 2.3 T (**Fig. 2A**), slightly lower than that of the bulk-form pristine sample (~ 2.8 T) (**Fig. 1A**). After charging at -1.2 V for one hour, the coercivity drastically decreased to ~ 1.0 T. We monitored the recovery of the coercivity during the discharging process by continuously recording the hysteresis loops. The coercivity increased monotonically with the discharging time (**Fig. 2A**), and regained most of its initial value in the early stage of the discharging process, reaching ~ 1.9 T within 10 hours (**inset in Fig. 2A**). After a prolonged time of discharging (~ 60 hours), the coercivity fully recovered.

In parallel, we studied the dynamics of the hydrogen charging/discharging process by observing the evolution of crystal structure with *in situ* X-ray diffraction (XRD) in transmission mode (**Fig. 2B**). Upon hydrogen-charging, all diffraction peaks were shifted to lower angles, and after one hour, the positions of the diffraction peaks stayed unchanged, indicating the complete charging of the whole sample. In the discharging process, two stages were discerned. First, strikingly, only a negligible shift of the peaks was observed over the first 10 hours (**Fig. 2C**), indicating that the bulk material is still charged with hydrogen atoms. Second, only after about 80 hours of discharging the peaks recovered to their original positions. This observation is in strong contrast with the substantial change in coercivity over the corresponding period (**Inset in Fig. 2A**). It suggests that the predominant coercivity change is not ascribed to the slow hydrogen desorption from the volume of the material.

Since the predominant change of coercivity does not arise from the volumetric slow diffusion of hydrogen atoms, we expect a relatively fast response of magnetization reversal to hydrogen charging (**Fig. 2C**). We first magnetized the as-prepared sample with 6.8 T (point ① in the inset). Then, the magnetic field was reversed to -1.1 T (point ②), smaller than the coercivity of the pristine sample (~ 2.3 T) and therefore, the magnetization remained positive and nearly constant. Upon hydrogen charging by applying -1.2 V (point ③), the magnetization decreased immediately and abruptly, and flipped from positive to negative in ~10 minutes. The magnetization started to level off after ~4 hours. The immediate response of magnetization reversal to the voltage stimulus confirms the existence of relatively fast diffusion path of hydrogen atoms.

**Atomic-scale tracking of hydrogen atoms within the hierarchical microstructure**

We carried out multi-scale multi-microscopy mapping of the micro- and nano-structural features to rationalize the fast diffusion pathways of hydrogen atoms. Optical microscopy (**Fig. 3A, Fig. 1**) showed that the sample was polycrystalline with grains of ~26 μm separated by high-angle grain boundaries (HAGBs). The grains were further divided into sub-grains by low-angle grain boundaries (LAGBs) as shown by electron back-scattered imaging (**Fig. 3B**). Inside the grain transmission electron microscopy shows the typical cellular structure, composed of the matrix cell with sizes of ~ 40 nm, the cell boundary and the Zr-rich lamellae crossing the cellular structure (TEM, **Fig. 3C, S4**). After tilting the c-axis of the specimen out of the viewing plane, high-resolution TEM and the corresponding selected area electron diffraction (SAED) showed that the matrix phase is $Sm_2Co_{17}$ (rhombohedral $Th_2Zn_{17}$ type) and the cell boundary phase $SmCo_5$ (hexagonal $CaCu_5$ type) (**Fig. 3D**). This hierarchical micro- and nanostructure matches previous reports[9-13].

We used atom probe tomography (APT) to locate hydrogen atoms and deuterium atoms within the hierarchical microstructure of the deuterium-charged samples. Isotopic marking by deuterium atoms (D) minimized the influence of residual hydrogen in the atom probe. We first analysed a specimen containing a HAGB (**Fig. S5, S6**). The element-specific atom maps in **Fig. 4A** reveal the Cu-rich cell boundaries, the matrix cells and the Zr-rich platelets, matching TEM results. Three-dimensional reconstruction shows that D mostly segregates in a 10–12 nm thick layer at HAGB, reaching a concentration of approx. 3.5 at% (**Fig. 4A**). A close face-on view (**Fig. 4B**) reveals that D segregates at the intersection of the cell boundaries with the HAGB. In the cells and cell boundaries, a very limited amount of D was detected within the structure close to the detection limit, with no noticeable partitioning difference between them (**Fig. S6**). In addition, deuterium atoms appear slightly depleted from the Zr-phase.

We then performed another analysis targeting LAGBs (**Fig. 4C, D, S7**). Again, deuterium appears depleted in the Z-phase, and no preferential segregation within cells and the cell boundaries (**Fig. S8**). Yet again, a strong segregation of deuterium at LAGB was observed. The corresponding top-view shows a series of linear features highlighted by a set of 0.35 at.% D iso-composition surfaces, which are likely the dislocations that constitute the LAGB[26]. The D-concentration at these dislocations can reach up to 0.4 at% D, and H up to 4–5 at%. Importantly, the cell edges are all connected to these dislocations and the cell structure stops abruptly at LAGB (**Fig. 4D**).

**Discussion**

The observation of hydrogen/deuterium segregation at GB regions (**Fig. 4**), coupled with *in situ* XRD, detecting no volumetric structural change (**Fig. 2B**), suggests that the substantial change of coercivity in the early stage of discharging process arises from the desorption of hydrogen atoms from GBs. The ability to modulate the coercivity by only charging/discharging GBs enables the fast control of coercivity, as verified by the immediate start of magnetization reversal upon hydrogen charging (**Fig. 2B**). The herein identified crucial role of grain boundary in controlling the coercivity explains why reducing the volume fraction of grain boundaries[27] or optimising the cellular structure near the grain boundary[28] can increase the coercivity of $Sm_2Co_{17}$-based magnets. Below we discuss how hydrogen segregation at GBs changes the coercivity, starting with the microstructural features near GBs.

Three microstructural features distinguish the GB region from the grain interior. First, compared with the continuous cellular structure in the grain interior (bottom part in **Fig. 4C**), the cell boundaries are broken and terminated near GBs (**Fig. 4A, C, D**). Second, the typical cell size and shape in the grain is approx. 40 nm and with a regular shape, but becomes larger and strongly elongated near GBs (**Fig. 4**). These results agree with recent TEM reports that the incomplete cellular structure near GBs extend towards the sub-micrometer scale[29,30]. Third, the composition profiles (**Fig. S6, S8**) show that the $SmCo_5$ phase contains almost twice as much Cu near the grain boundary, i.e. 30 at.% compared with 15 at.% in the grain interior.

The observed different cellular structure and microchemistry near GBs significantly reduces the local nucleation field required for magnetization reversal. According to the micromagnetic theory, the critical nucleation field, $H_n$, can be described by[14]

$$H_n = \frac{1}{2M_s\Delta}\left(\gamma_{SmCo_5} - \gamma_{GB}\right) - DM_s \qquad (1)$$

in which $\Delta$ is the width of the transition region where the domain-wall energy changes from $\gamma_{GB}$ in the grain boundary to $\gamma_{SmCo_5}$ in the cell boundary, and $D$ is the demagnetizing factor. The $SmCo_5$ phase has much larger magnetocrystalline anisotropy than GBs, and determines the domain-wall energy difference, $\left(\gamma_{SmCo_5} - \gamma_{GB}\right)$. Near GBs the Cu concentration in the $SmCo_5$ phase becomes twice that in the grain interior, which substantially reduces its magnetocrystalline anisotropy[31,32] and thus the domain wall energy, $\gamma_{SmCo_5}$. In addition, the disrupted $SmCo_5$ phase allows the easy movement of domain walls through the matrix phase, triggering a macroscopic magnetization reversal. These account for the preferential nucleation of reversed domains near GBs (**Fig. 1**). Moreover, when the $SmCo_5$ phase was charged with hydrogen atoms, its magnetocrystalline anisotropy will decrease by ~40%[20]. This further decreases the domain-wall energy of the $SmCo_5$ phase and the nucleation field. Besides, hydrogen segregation may enlarge the transition region ($\Delta$) between GBs and the $SmCo_5$ phase with its continuous concentration change and reduce the nucleation field. Hence, hydrogen segregation acts here as a tool to further weaken the nucleation field of the GB region and amplify its effect in initiating the magnetization reversal. Next, we consider the mechanism behind the propagation of the initially-nucleated magnetic domains near GBs into the grain interior.

In the grain interior, the continuous network of the SmCo$_5$ phase subdivides the individual grains of the Sm$_2$Co$_{17}$ matrix into a nanoscale cellular structure, rendering them into classical pinning-type magnets. However, the models to explain their magnetization reversal assume that grain boundaries should be non-ferromagnetic and one-to-two atomic layers thick to reduce the associated stray field negligibly[14,33]. This is not the case in the current material because of the expanded region near GBs with the disintegrated cellular structure and different microchemistry. We can describe the demagnetization process of the whole grain triggered by the initial demagnetization near GB as follows. The local magnetic field is a superposition of the external field $H_{ext}$ and the local demagnetization field, $N'M$, where $N'$ is the local or effective demagnetization factor and $M$ the net magnetization of the sample. The latter can be significantly inhomogeneous, and reach values much larger than the net demagnetization field[3], $H_D = NM$, $N$ being the demagnetization factor of the sample (*3*). As discussed earlier, the nucleation field of the GB region, $H_{c,GB}$, is much smaller than $H_c$ of the grain interior, $H_{c,g}$. Under a small external field, we have $H_{c,g} > H_{c,GB} > H_{ext} + NM$. As the local magnetic field, $H_{ext} + NM$, approaches $H_{c,GB}$, the initial nucleation of magnetic domains occurs near GBs (**Fig. 1**), producing a local demagnetization field, $N'M_s$, where $M_s$ is spontaneous magnetization of the main Sm$_2$Co$_{17}$ phase. Then, the adjacent inner layer with higher coercivity, $H_{c,g}$, is under the higher magnetic field, $H_{ext} + NM + N'M_s$. This additional negative field, $N'M_s$, can be of 0.5-1.0 T, depending on microstructural features and $M_s$[34]. Thus, if $H_{c,g} - H_{c,GB} < N'M_s$, the demagnetization of GB region will inevitably trigger an avalanche-like demagnetization process in the whole grain, driven by the local enhancement of the demagnetization field.

**Conclusion**

In summary, our results show that by electrochemically-controlled hydrogen charging/discharging the coercivity of the pinning-type Sm2Co17-based magnet can be tuned by an unprecedented value of ~ 1.3 T, the highest value ever reported by a magneto-electric approach. In situ magneto-structural characterization and atomic-scale tracking of hydrogen atoms over the hierarchical microstructure reveals that the predominant change of the coercivity arises from the decoration of the grain boundaries with hydrogen atoms. These findings reveal, contrary to the conventional paradigm, a critical role of the grain boundaries in determining the coercivity in pinning-type magnets. Furthermore, these discoveries are anticipated to apply to other technologically important pinning-type magnets such as FePt and MnAl. Future performance-optimisation strategies should consider engineering of grain boundaries to enhance the coercivity of pinning-type magnets for the clean-energy applications. The demonstrated voltage-controlled and giant modification of the coercivity by hydrogen insertion (extraction) into grain boundaries also opens a way to various applications such as magneto-electric actuation or sensing in which large magneto-electric effect are needed.

**Materials and methods**

**Materials and microstructure characterization.** The $Sm_2Co_{17}$-type permanent magnets with dimensions of Φ10 mm×6 mm were purchased from Sigma-Aldrich (Stock No. 692832). The composition of the powder was analyzed by inductively-coupled plasma mass spectroscopy (**Table S1**) and its microstructure characterized by optical microscopy (KIT), powder X-ray diffraction with a Mo $K_{\alpha,\beta}$ source (Philips X'Pert Analysis, KIT), field-emission scanning electron microscope (SEM) equipped with energy dispersive X-ray spectroscopy and electron channeling contrast imaging (ECCI) (Zeiss Ultra 600/Merlin, both at KIT and MPIE), and transmission electron microscope (TEM, FEI Titan 80-300, KIT). Before optical and SEM characterization, the sample surface was mechanically polished. The preparation of TEM samples followed the ordinary procedure of cutting, lifting and milling using FIB/SEM dual beam system (FEI Strata 400 and Zeiss Auriga 60, KIT). The TEM observations were taken both with the c-axis of the crystal structure out of and parallel with the viewing planes.

**Preparation of the $Sm_2Co_{17}$ powder electrode and the electrochemical set-up.** To prepare the $Sm_2Co_{17}$ electrode, the as-received bulk sample was first charged with hydrogen atoms. The insertion of hydrogen atoms caused the expansion of the sample, and, consequently, the surface of the bulk sample collapsed into the very large particles. These particles were mixed with PVDF solution to form a slurry, which was then coated onto thin copper foils (thickness ~ 15 μm). The slurry/Cu foil composite was then put into a homogeneous magnetic field to magnetically align the particles. Afterwards, it was further dried at room temperature for overnight. As the last step, the composite was compressed under a pressure of ~ 100 MPa to further fix the particles and to increase the electrical conductivity between $Sm_2Co_{17}$ particles and the Cu foil. We prepared the PVDF solution by dissolving PVDF powder in NMP solution at a mass ratio of 5:95 with stirring overnight.

The charging and discharging of the $Sm_2Co_{17}$ electrodes were carried out under potentiostatic control in a three-electrode electrochemical system (Autolab PGSTAT 302N, KIT). The working, the counter and the reference electrodes were the $Sm_2Co_{17}$ powder electrode, Pt wires and a pseudo Ag/AgCl electrode, respectively; the electrolyte was an aqueous electrolyte of 1 M KOH prepared from ultrapure water with a resistivity of ~ 18.2 MΩ (**Fig. S2**). The potential of the peuso Ag/AgCl electrode is 0.300±0.002 V more positive than the standard Hg/HgO (1M KOH) electrode, and for comparison, all the voltages in the paper were converted to the Hg/HgO scale. According to the cyclic voltammogram of the $SmCo_{7.35}$ electrode (**Fig. S3**), the voltages steps of -1.2V and -0.4 V were used to charge and discharge the sample, respectively.

***In situ* XRD measurement.** The crystal structure of the $Sm_2Co_{17}$ electrode under the application of -1.2 V and -0.4 V was monitored by *in situ* XRD with a parallel beam laboratory rotating anode diffractometer (Mo $K_\alpha$ radiation) in transmission geometry. The transmission geometry allowed the detection of the entire volume of the $SmCo_5$ particles rather than only their surfaces. For *in situ* measurement, the $Sm_2Co_{17}$ electrode was attached to a glass plate (thickness ~ 0.1 mm) and then immersed in the 1 M KOH electrolyte in plastic bags. The counter and reference electrodes were the Pt wire and the pseudo Ag/AgCl electrode, respectively. Diffraction patterns were collected every 10 minutes with a Pilatus 300K-W area detector. NIST SRM660b $LaB_6$ powder was used for the detector calibration and the determination of the instrumental resolution.

***In situ* SQUID measurement.** *In situ* magnetic measurement was carried out with a custom-built miniaturized Teflon electrochemical cell in a superconducting quantum interference device (SQUID, MPMS3, KIT) at room temperature. In the electrochemical cell, the $Sm_2Co_{17}$ electrode, Pt foil and peuso Ag/AgCl electrode were the working, counter and reference electrodes, respectively. The electrolyte was 1 M KOH. The $Sm_2Co_{17}$ electrode and the Pt foil were attached to the flat surface of a plastic rod, and the reference electrode was threading through a capillary to determine the applied potential of the working electrode. The magnetic measurements were performed at the sealed mode of SQUID and with the applied magnetic field parallel to the surface of the Cu foil. The magnetic hysteresis loop of the bulk sample was measured with the magnetic field along the c-axis of the particles.

**MOKE measurement**

The magnetization reversal process was monitored by characterizing the magnetic domain structure under the external magnetic field using magneto-optical Kerr effect (MOKE) microscopy (Zeiss Axio Imager, D2m evico magnetics GmbH, TU Darmstadt). Before the MOKE observation, the sample was fully magnetized at a pulsed field of 6.8 T. Then, the magnetic domain structure was observed after applying a reversed magnetic field of 1.5 T, 2 T, 2.2 T, 2.5 T, 2.8 T, 3.0 T and 6.8 T. The magnetic field was applied parallel to c-axis of the matrix $Sm_2Co_{17}$ phase, and the images taken with the c-axis in the viewing plane. To enhance the image contrast, the non-magnetic background image was subtracted from the collected average image using KerrLab software.

**APT measurement**

For the atom probe tomography (APT, MPIE) measurement of hydrogen distribution, the $Sm_2Co_{17}$ electrode was charged at -1.2 V for ~2.5 hours in 0.1 M NaOD in $D_2O$ (instead of $H_2O$) using the three-electrode system as described above. After the full charging, the sample was cleaned by ethanol and was transferred in 10 minutes to FIB chamber for the cutting, milling and lifting at room temperature (FEI Helios Nanolab 600/600i). 3-4 APT tips were prepared within 3-4 hours at room temperature. Annular milling was used to sharpen the needle-shaped morphology with a diameter less than ~100 nm. After that, a cleaning of the specimen at 5 kV was performed to remove the beam-damaged surface regions. The prepared tips were transferred into the load-lock chamber of APT, waited ~ 2 hours until the vacuum reached ~$10^{-8}$ Pa (LEAP 3000 XHR), and then transferred into analysis chamber (~$10^{-11}$ Pa) at 70 K. We also acquired data on Cameca LEAP 5000XR, and only waited for half an hour before transferring the sample from APT load-lock to analysis chamber. The APT experiments were conducted using high-voltage mode with a pulse fraction of 15% at a base temperature of 70K, a pulse frequence of 200 kHz and an evaporation rate of 0.5. Atom probe data reconstruction and analysis were performed by CAMECA IVAS 3.8.4 software.

**Acknowledgements**
The authors appreciate financial support from Deutsche Forschungsgemeinschaft under contract number HA 1344/34-1 (R.K, H.H.) and CRC/TRR 270, (L.S., K.S., O.G., BG) as well as Alexander von Humboldt Foundation (X.Y.). BG and LTS acknowledges financial support from the ERC-CoG-SHINE-771602. **Author contributions:** X.Y., R.K., H.H. conceived the project. X.Y. designed the experiments. X.Y. conducted material preparation and microstructural observation, built in situ electrochemical charging/discharging setup and conducted magnetic measurements; F.Y. conducted APT, part of SEM, assisted by L.T.S, and F.Y., B.G. processed APT data; L.S., X.Y. performed MOKE; H.G., X.Y. performed in-situ XRD; M.R.C. prepared TEM samples, discussed results, and D.W, W.W conducted TEM observation. K.S. and O.G contribute significantly to the interpretation of results. X.Y. wrote the initial draft and B.G., K.S., R.K. revised the manuscript. All authors participated in the discussions, contributed to improving the manuscript and approved the submitted manuscript. **Competing interests:** None. **Data and materials availability:** All data needed to evaluate the conclusions of the study are present in the paper or the supplementary materials.

**Supplementary Materials:**
Fig. S1 to S8
Table S1

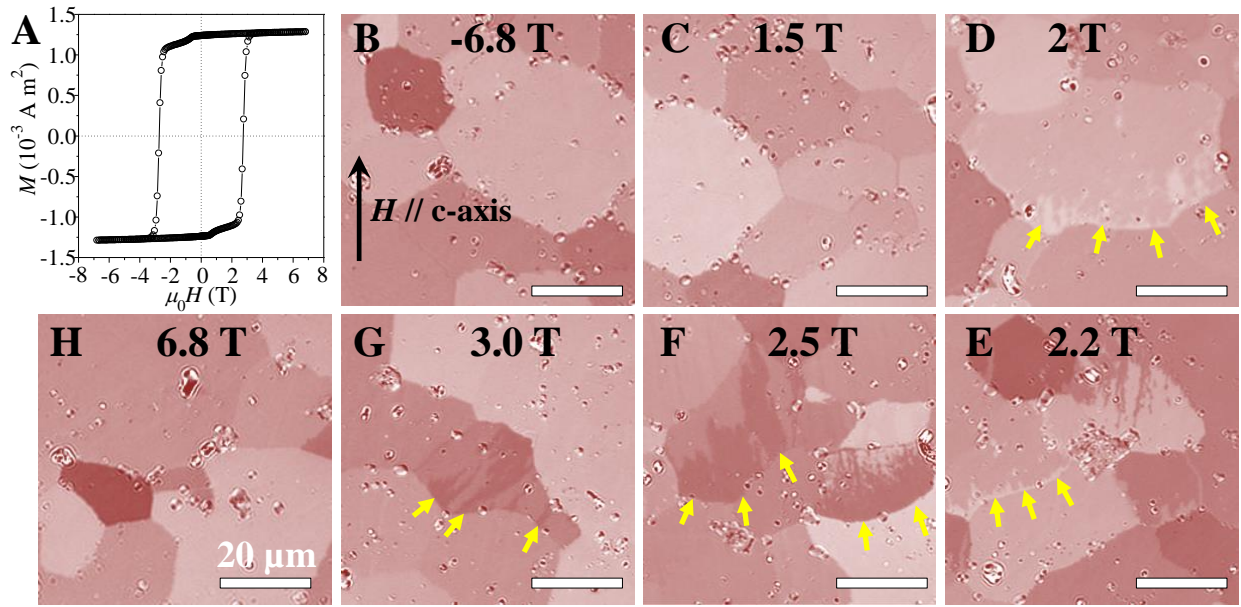

**Fig. 1 Magneto-optical Kerr effect (MOKE) microscopy observation of the magnetization reversal process in Sm$_2$Co$_{17}$-based magnet** (**A**) Hysteresis loops of the bulk Sm$_2$Co$_{17}$-type magnet. (**B-H**) MOKE observation of the magnetic domain structure under different demagnetization field from the magnetically-saturated state – the applied magnetic field is parallel to c-axis of crystal structure as indicated in (**B**).

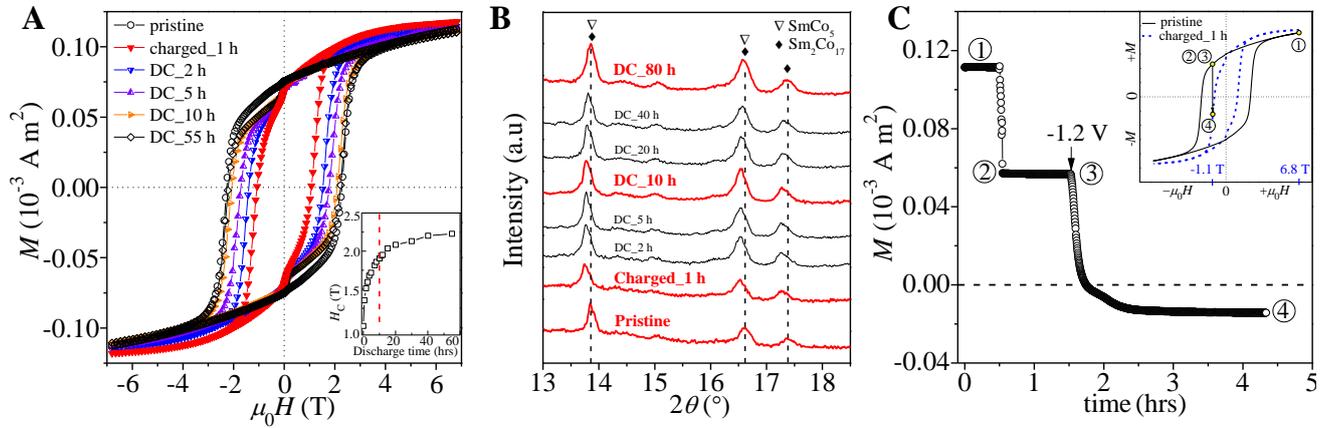

**Fig. 2. Reversible modification of the coercivity in Sm$_2$Co$_{17}$-based magnet by hydrogen charging/discharging and its dynamics.** (**A**) Hysteresis loops of the as-prepared sample and those after hydrogen charging at -1.2 V for 1 hour (charged_1h) and after further discharging for 2 h (DC_2h), 5 h (DC_5 h), 10 h (DC_10 h) and 55 h (DC_55 h). **Inset** shows the continuous change of the coercivity against the discharging time. (**B**) *In situ* XRD monitoring the evolution of the crystal structure of the as-prepared sample during charging and discharging, showing that within the first 10 hours of discharging the diffraction pattern remained nearly unchanged compared with the charged sample. (**C**) Time evolution of the magnetization in the as-prepared sample with the voltages switched to -1.2 V, showing the immediate response of magnetization reversal to voltage stimulus. Points ①, ②, ③, and ④ indicate different magnetization states as shown in the **inset**.

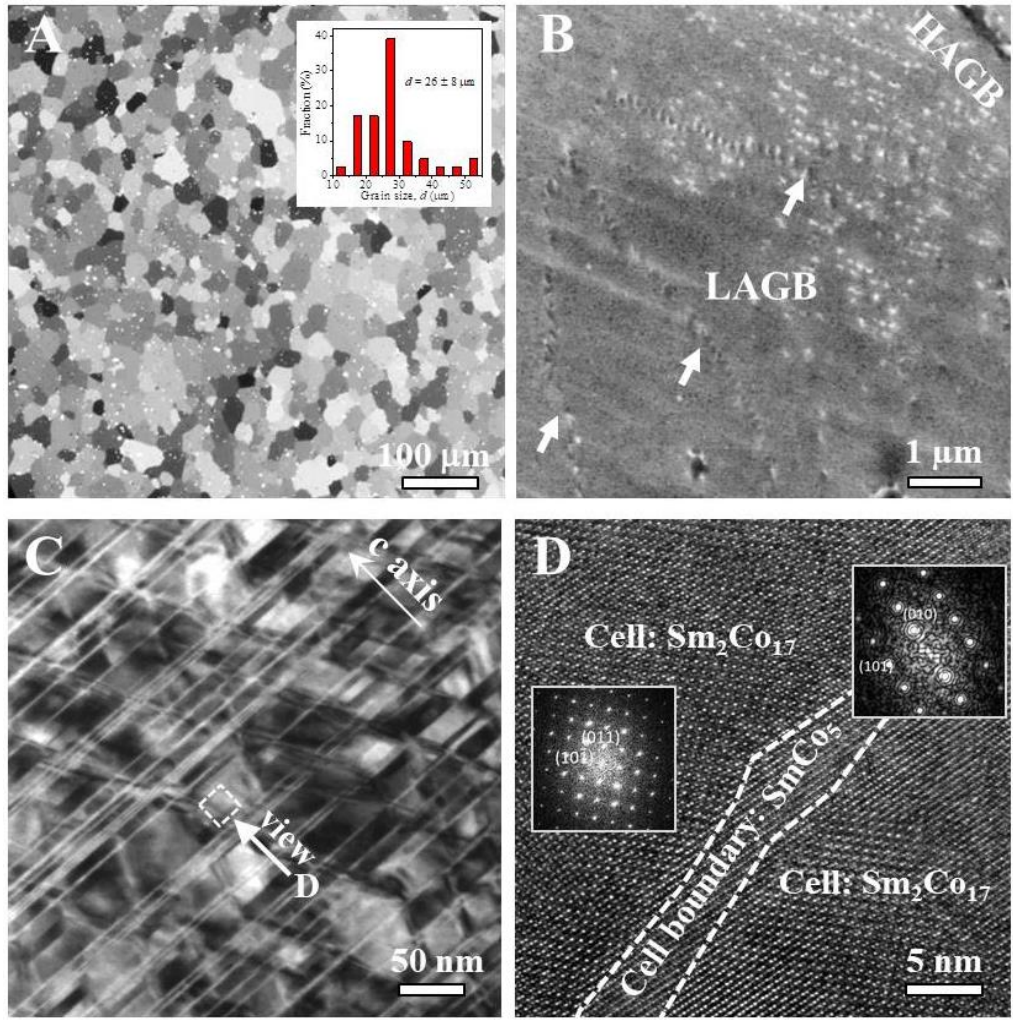

**Fig.3. Multi-level multi-spectroscopy characterization of the hierarchical microstructure in Sm$_2$Co$_{17}$-based magnet.** (**A**) An optical microscopy image showing the polycrystalline grains with grain sizes ~ 26 μm. (**B**) An enlarged image of the crystal grain showing low-angle grain boundaries (LAGB) within the grain. (**C**) A bright-field TEM image within the grain showing the nano-scaled cellular structure, composed of the Sm$_2$Co$_{17}$ cell, the SmCo$_5$ cell boundary and the Zr-rich platelets. (**D**) A close-up on the cell and cell boundary by HRTEM and their Fourier transformation. The c-axis of the matrix phase in (**C**) and (**D**) is in and out of the viewing plane, respectively.

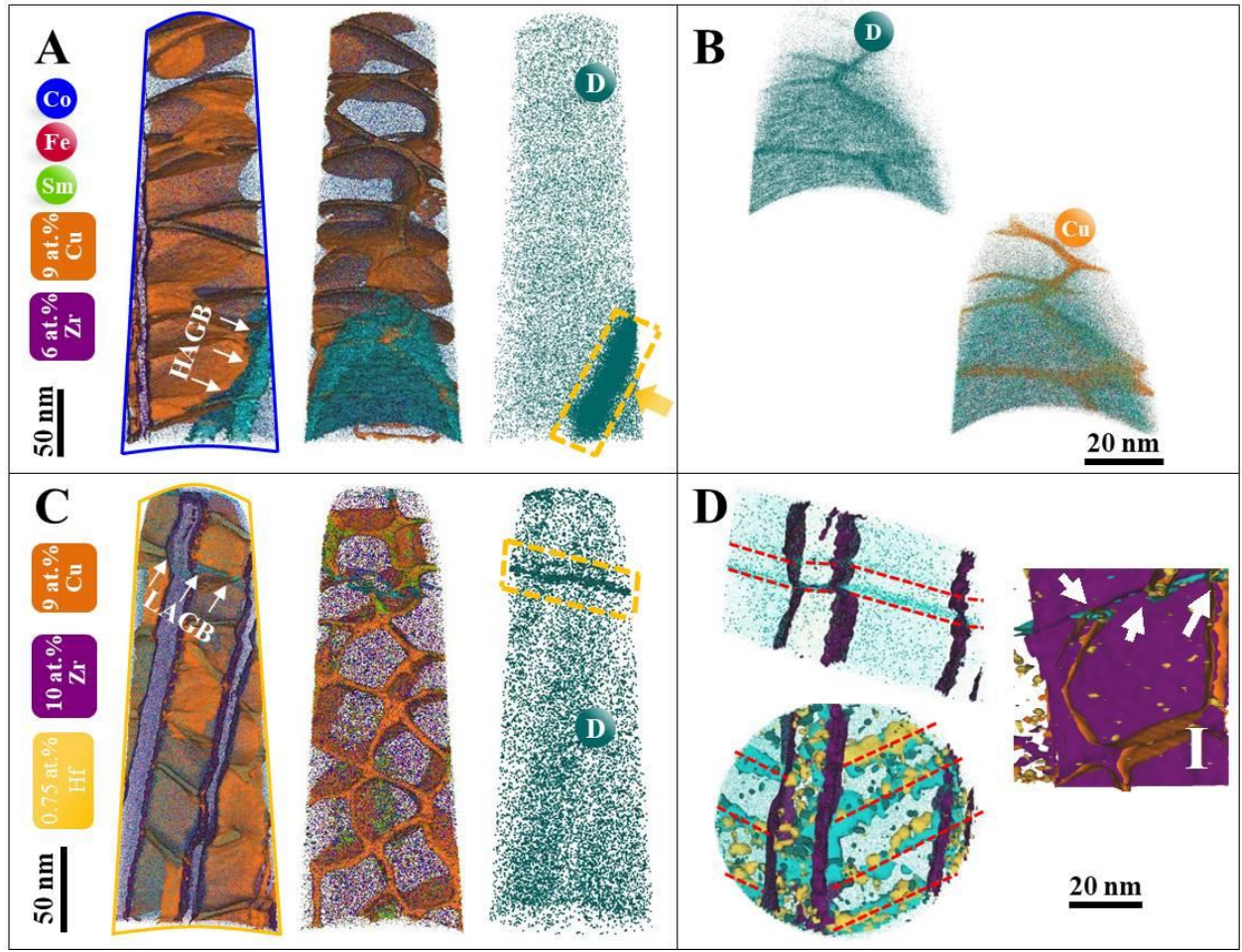

**Fig. 4 Atomic-scale tracking of hydrogen/deuterium atoms within the microstructure in Sm$_2$Co$_{17}$-based magnet.** (**A**) 3D atom map of a deuterium-charged sample containing a high-angle grain boundary (HAGB). The sets of 9 at.% Cu and 6 at.% Zr iso-composition surfaces evidence the Cu-rich cell boundaries and the Z-phase; the distribution of deuterium reveals strong planar segregation near HAGB. (**B**) A face-on view on the distribution of D and Cu using 3.5 at.% D and 9 at.% Cu near the HAGB highlighting the co-location of D and Cu. Note the incomplete cellular structure and its abrupt stop at HAGB. (**C**) 3D atom map of a deuterium-charged specimen containing a low-angle grain boundary (LAGB). Note the disruption of the cellular structure adjacent to LAGB. The distribution of D shows the strong planar segregation at LAGB, the depletion at Zr-rich phase and no partioning between cell and cell boundary. (**D**) A close-up using 0.35 at% D, 0.75 at% Hf and 4.5 at% H iso-surfaces showing D segregation along a series of linear features interpreted as dislocations, each of which is connected to a cell edge; a close-up on the LAGB showing the direction connection of each dislocation to a cell boundary as indicated by white arrows (right side in **D**).